# IMPROVE KNOWLEDGE OF ACOUSTIC FACTORS INVOLVED IN RAILWAY NOISE ANNOYANCE

**M. Sineau[1*], F. Mietlicki[1], D. Bernfeld[1], S. Cherfan[1], L. Motio-Betato[1]**
**M. Hellot[1], A-S. Evrard[2], J-P. Regairaz[3], C. Rosin[3]**
[1] Bruitparif, 32 boulevard Ornano, 93200 Saint-Denis, France
[2] Université Lyon 1, Université Gustave Eiffel, Ifsttar, Umrestte, UMR T_9405, Bron, France
[3] SNCF-Réseau, 10 rue Camille Moke, 93210 Saint-Denis, France

## ABSTRACT

Energy noise indicators are generally used to characterize the exposure of populations to transportation noise in relation to their long-term annoyance, but they do not adequately reflect the repetitive nature of noise peaks generated by railway traffic. The GENIFER project aims to test a study protocol designed to rank railway noise events according to the instantaneous annoyance they cause to residents. This study will be carried out in a sector exposed to railway noise in the Île-de-France region and will require the recruitment of 60 volunteer local residents. It will propose the use of innovative tools for collecting information, including an electronic remote-control allowing participants to rate the annoyance they feel when trains pass by, and noise sensor instrumentation allowing the simultaneous collection of the acoustic characteristics of railway noise peaks. It also includes semi-directive interviews and a questionnaire to identify co-determinants of annoyance. The instantaneous annoyance scores collected will also be compared with those obtained from commented listening to sound samples of passing trains. In addition to assessing the acceptability of the protocol by the participants, this study aims to validate the feasibility of ranking railway noise events according to their acoustic characteristics in terms of the annoyance expressed.

## 1. CONTEXT

Rail transport has many environmental advantages in terms of decarbonizing the economy and travel. It is also an appropriate response to the challenges of mass mobility for passengers and freight. However, the noise generated by rail traffic impacts local populations and represents a major negative externality of this mode of transport, which can sometimes even be a brake on its development. According to the study on the social cost of noise in France published by Ademe (Agency for Environment and Energy Management) and the Conseil national du bruit (Noise National Council), rail traffic noise represents a cost to French society of 11.1 billion euros, i.e. 7.6% of all the social costs generated by noise in all its dimensions (transport, neighborhood, work) in France. At the national level, nearly 1.4 million people are said to be very seriously affected by noise from rail traffic and nearly 700,000 people suffer from sleep disturbance [1]. The health impacts of rail noise, even if it affects fewer people than road noise, are now well established scientifically, particularly in terms of long-term annoyance and sleep disturbance, and have been strongly recommended by the WHO [2]. However, it appears that the instantaneous annoyance that may exist more specifically in relation to railway noise is much less well documented. The acoustic energy indicators used to characterize exposure to railway noise (which correspond to noise "averages" over given periods of time) do not meet the expectations of the exposed populations, particularly as they do not adequately reflect the repetitive nature of railway noise and the instantaneous annoyance of passing traffic. The discussions conducted in France by the Ministry of Ecological Transition and the Conseil National du Bruit on these subjects have revealed the interest of introducing into the regulations, in addition to energy indicators, event-based indicators and railway event counting [3]. In order to be able to count railway events in a relevant manner, it would seem necessary to be able to associate a weighting to each one, which could take into account the degree of instantaneous annoyance that it is likely to generate, being understood that certain trains (such as freight trains or High Speed Trains) seem to be particularly annoying due to some of their acoustic characteristics (intensity, duration, emergence, spectral content, suddenness, etc.) and/or their period of appearance (times of passage) [4]. This requires the ability, for a given site, to rank rail traffic according to the degree of instantaneous annoyance it generates for local residents, and to determine which descriptor and/or combination of acoustic descriptors best correlate with this instantaneous annoyance. The GENIFER project (improving knowledge of the acoustic factors of the instantaneous annoyance due to railway noise), implemented by Bruitparif, the Gustave Eiffel University and SNCF Réseau, is part of this framework and proposes to carry out a feasibility study seeking to better understand the acoustic factors involved in the instantaneous annoyance expressed by residents with regard to noise generated by railway traffic. It is an essential preliminary step if we want to be able to develop and introduce into the regulations relevant indicators for characterising railway events.

This project is supported by the French National Research Program for Environmental and Occupational Health (ANSES-22-EST-182) of ANSES (French Agency for Food, Environmental and Occupational Health & Safety. The study started in November 2022. This article presents the proposed methodology for its implementation. The field survey is planned to be carried out from April to June 2023.

## 2. OBJECTIVES OF THE STUDY

The main objective of GENIFER is to evaluate the feasibility of a study seeking to better understand the acoustic factors involved in the annoyance expressed by local residents with regard to noise generated by rail traffic. It proposes to develop and test a study protocol designed to classify and rank railway noise events according to their associated instantaneous annoyance and their acoustic characteristics. It will validate the interest of carrying out a study on a larger scale to meet the following specific objectives.

### 2.1 Objective 1: Construction of a survey methodology, selection of a study site and development of tools for data collection

The first objective of this feasibility study is to propose a survey methodology and to select a pilot site. It is also to propose tools for collecting information, in particular an interview grid and a questionnaire. Finally, an electronic device in the form of a remote control, specially developed by Bruitparif for the project, will enable participants to report their instantaneous annoyance when trains pass by.

### 2.2 Objective 2: Test the implementation of the survey methodology and data collection tools in the field

The second objective is to implement the field survey on a pilot site in Île-de-France through concomitant data collection:

- Acoustic data through noise measurements to determine the values of the different acoustic descriptors associated with each rail traffic during the whole survey period and through numerical modelling of rail noise on the pilot site.
- Participants' responses through semi-structured interviews, questionnaires, recording of instantaneous annoyance when trains pass by using an electronic remote control, and commented listening in participants' homes.

This implementation in real-life situations will make it possible to assess the relevance of the study protocol, to check its acceptability by the participants and to compare the different approaches.

### 2.3 Objective 3: Classification and ranking of railway noise events

The third objective is to analyze the noise measurements and to exploit the results of the semi-structured interviews, the questionnaires, the collection of instantaneous annoyance ratings for passing trains and commented listening. The railway noise events will be classified according to their instantaneous annoyance rating and their acoustic characteristics.

## 3. ORGINALITY OF THE PROJECT

The originality of GENIFER lies in three aspects.

- Assessment by residents living near railway lines, at their homes, of the instantaneous annoyance due to the noise of train passages by rating this annoyance at each train passage by means of an electronic remote control. The annoyance ratings will be related to the acoustic characteristics of the train passages, determined by noise measurements and modelling carried out in the vicinity of these residents. This approach has never been carried out in France, in situ, to assess the instantaneous annoyance due to railway noise.
- The comparison of the results of the questionnaires and the annoyance scores collected by the remote controls with the railway events and the pass-by noise measurements will make it possible to classify the trains by annoyance score and then to group them according to their acoustic characteristics. This will make it possible to determine the acoustic descriptors or combinations of acoustic descriptors that appear to be most relevant in explaining the annoyance expressed.
- This pilot survey will also allow us to compare different approaches to collecting participants' feelings: semi-structured interviews, rating of instantaneous annoyance using a remote control and commented listening to sound samples recorded at the pilot site.

This project will make it possible to verify the feasibility of collecting information directly from railway residents chronically exposed to railway noise on the annoyance felt when trains pass by, and to relate this annoyance to the acoustic characteristics of train passages evaluated simultaneously in the environment. It thus differs from psychoacoustic studies of short-term annoyance, which are carried out in the laboratory with a panel of volunteer participants who are asked to judge the noise in different situations in which they are artificially immersed. It also differs from the usual surveys of long-term noise annoyance.

## 4. DESCRIPTION AND IMPLEMENTATION

### 4.1 Survey tools

The first step was to establish the main methodological guidelines for the conduct of the survey and to develop the evaluation tools: interview grid, questionnaire, remote control for train scoring.

*4.1.1 Semi-structured interview grid and questionnaire*

The first tool developed is a guide for conducting semi-structured interviews with about ten participants in the study. The semi-structured interview, based on themes and using open-ended questions, enables the interviewees to talk about their experiences, behavioral aspects and the cognitive factors (beliefs/knowledge about the effects, intention to act, attitudes towards the noise source, evaluation of the costs) that modulate them. They will take place face-to-face in the homes of residents of the pilot site. An interview grid has been drawn up based on elements from the literature. The targeted duration of this interview is around 30 minutes. This interview will allow the participant to express himself spontaneously, and in his own words, on his feelings about railway noise. The results of these interviews will be used to develop a more direct questionnaire for a nationwide study. The general questionnaire will be designed to collect information on the participants: demographic and socio-economic characteristics, the characteristics of their dwelling (type of housing, insulation, etc.), their overall perception of their environment, their perception of transport, particularly rail transport, and their personal and professional situation. This questionnaire was inspired by similar work, in particular the DEBATS study [5], which aim was to gain a better understanding and quantify the effects of aircraft noise on the health of people living near airports. The questionnaire will take about 45 minutes to complete.

*4.1.2 Remote control for the marking of passing trains*

A central part of the survey is for participants to record the levels of instantaneous annoyance experienced during different train passes of different categories using a connected remote control. The precise time of the train passage at the time of scoring is an essential information to associate this score with the characteristics of the railway events measured in the environment. First experiments conducted by Bruitparif using paper questionnaires showed that it was difficult for participants to indicate precisely the time of passage and that it was then tedious or even impossible to associate the annoyance scores with the corresponding trains and the measured noise levels. For this reason, an electronic remote control was developed. The remote control will record the exact date and time of the passage of the evaluated train, an overall annoyance score due to train noise on a scale of 1 (lowest annoyance) to 10 (highest annoyance) as well as the state of the windows (open/closed) or whether the participant was outside at the time of the scoring. This remote control will be connected to the Bruitparif server in order to store the information in real time.

### 4.2 Selection of the pilot site

The study is carried out on a pilot site in the Ile-de-France region. A Geographic Information System was used to select the site. It was used to cross-reference different data sources and filter them to select only those sectors that met the following selection criteria:

- The proximity of the rail network.
- The presence of populations exposed to railway noise levels above 54 dB(A) in Lden (WHO quality objective) and 73 dB(A) in Lden corresponding to the limit value retained by the regulations to define railway noise black spots on conventional tracks. The strategic noise maps drawn up by numerical modelling in the framework of the European directive 2002/49/EC were used to identify these sectors,
- A number of daily train movements between 250 and 500 trains per day. This value is a compromise that makes it possible to avoid a low traffic situation that would make it very difficult to rate several consecutive train passages or, on the contrary, to avoid a situation with too much traffic that would make it difficult to rate and analyse the acoustic measurements for very close or simultaneous train passages.
- Various types of traffic: site with different types of traffic like regional trains, mainline trains, freight trains with night-time traffic.
- A high population density to facilitate the recruitment of participants and the possibility of having three groups of participants according to their exposure to railway noise (moderate, intermediate and high).
- A diversity of housing types (collective, individual).
- An area with no or low exposure to other noise sources. The strategic noise maps drawn up in the framework of the European directive 2002/49/EC will be used to exclude areas that are highly exposed to road and air traffic noise.
- The distance from railway stations, to exclude areas too close to stations where trains are likely to stop and to avoid the influence of sound announcements in stations.
- Sections without noise barriers or other source protection, to keep railway noise exposure homogeneous according to the floor of the participants.

The selected pilot site meets all these criteria and is located in the commune of Savigny-sur-Orge in the south of the Île de France region. The selected area includes more than 1300 dwellings. It offers a sufficient potential number of dwellings to recruit about 60 participants.

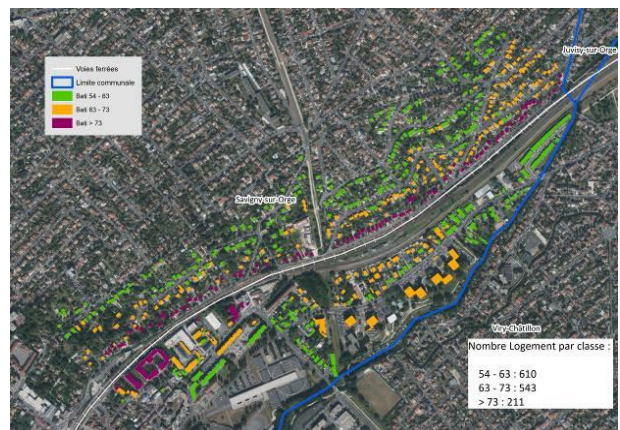

**Figure 1**. View of the pilot site

### 4.3 Recruitment of participants

In order to have participants with a contrasting exposure to railway noise, the size of the study area around the pilot site was designed to allow for the recruitment of 60 participants spread over three railway noise zones (20 participants per zone). This total number of participants is a compromise that allows for enough people per noise zone while remaining logistically manageable.

| 20 people | 20 people | 20 people |
|---|---|---|
| Group High exposure | Group Intermediate exposure | Moderate exposure group |
| Lden rail ≥ 73 dB(A) | 63 ≤ Lden rail <73 dB(A) | 54 ≤ Lden rail < 63 dB(A) |

*Lden 73 dB(A) = Regulatory limit value (Noise Black Spot)*
*Lden 54 dB(A) = WHO quality objective [2]*

Information was provided by the Municipality of Savigny-sur-Orge, which relayed it on its website and on social networks. An information letter was distributed to all the inhabitants of the study area. Door-to-door canvassing was also carried out. These actions resulted in the identification of 60 volunteers to participate in the study, including at least 20 people in each of the three groups of exposure to railway noise.

### 4.4 Pilot site instrumentation

*4.4.1 Noise measurements*

In order to assess the railway noise exposure of the participants, about 15 noise measurement systems were installed in the environment simultaneously with the survey: 14 classical stand-alone sonometer boxes placed all over the study area and an expert system (medusa), placed in the centre of the study area, developed by Bruitparif and having an automatic detection functionality of the direction of sound origin. This functionality allows the detection of sound events coming from a specific area. It will thus be possible to detect all rail traffic in an exhaustive manner. These measurement systems will provide Leq,100 milliseconds noise levels for the overall A and C weighted levels.

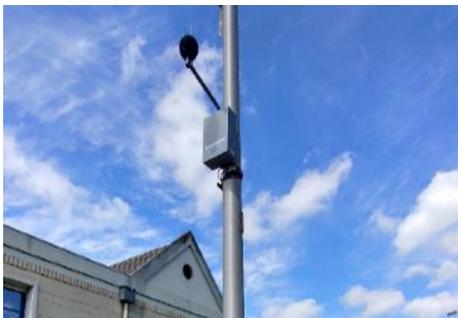

**Figure 2**. Temporary noise measurement system (Classic sound level meter)

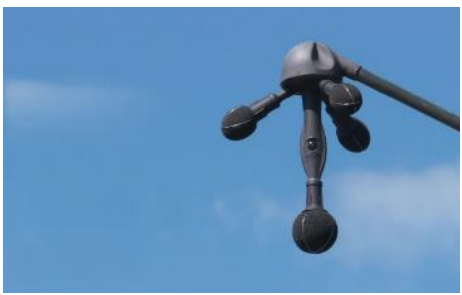

**Figure 3**. Expert system (Medusa) for noise measurement with direction of sound detection functionality

*4.4.2 Numerical modelling of railway noise*

A fine numerical modelling of railway noise will be carried out on the entire pilot site in order to simulate the sound levels generated by the passage of a train at any point in the territory and to determine the corrections to be applied to the acoustic descriptors measured at the measuring stations. An evaluation of the acoustic descriptors at the facades of the dwellings of the various participants can thus be obtained. This calculation model will be validated by means of noise measurements carried out within the framework of the study.

*4.4.3 Digital audio recordings*

Digital audio recordings will be made in order to build up a database of sound samples representative of the main families of trains running at the pilot site. After selection, following the analysis of the noise measurements, 3 to 5 sound samples will be retained per category of rolling stock, i.e. approximately 15 to 20 samples in total. The duration of the samples will be equivalent to the duration of the trains' passage, i.e. from a few seconds to about two minutes.
The digital audio recordings made will allow the participants to listen to different types of trains and to better understand their feelings and perceptions of the acoustic characteristics of train passages.
The recordings will be made using a binaural system that allows sound recordings to be made close to the human auditory system. The listening will then be done with headphones on systems calibrated to faithfully reproduce the recorded sounds and immerse the listener in the scene. The sound recording will be carried out in a location that is not disturbed by other sound sources, with a direct view of the tracks at an angle of approximately 120° and at a distance of between 20 and 50 meters from the tracks.

### 4.5 Conducting the survey

*4.5.1 Interviews and questionnaires*

Semi-structured interviews will be carried out with about ten participants. The results of these interviews will be used to refine the general questionnaire, particularly with a view to a larger-scale study. The other 50 participants will answer a closed questionnaire to collect information on their housing, their appreciation of the neighborhood, noise in general and railway noise in particular.

*4.5.2 Rating of disturbance to passing trains*

One of the central components of the survey is to provide participants with an electronic device for rating the instantaneous annoyance of passing trains. This device takes the form of an electronic box the size of a remote control. The device allows participants to assign a score of annoyance (from 1 to 10) due to the noise of a particular passing train. Each participant will have to carry out a total of three hours of train scoring in 6 sessions of 30 minutes or 3 sessions of one hour. These sessions will be carried out during preferential time slots determined in consultation with the Bruitparif interviewers and according to the participant's lifestyle. The sessions will be dedicated only to the rating of trains and not to other activities, so that the annoyance rating is not influenced by the type of activity. The participant will be able to specify the scoring conditions: inside the home with the windows closed or open or outside the home. This electronic scoring device will allow each train passage to be precisely time-stamped with the corresponding annoyance score. These annoyance ratings can then be associated with the acoustic characteristics of railway events measured in the environment.

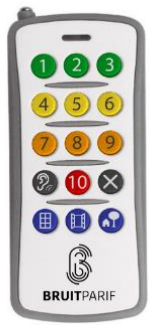

**Figure 4**. Electronic device for rating instantaneous annoyance when a train passes.

*4.5.3 Listening with commentary*

In a second phase, after the noise data has been analyzed, commented listening sessions will be carried out in the homes of 30 participants (half of all participants) selected from the 60 previously interviewed, so as to have ten people in each of the three railway noise exposure zones. The recording method used is to reproduce natural human sound perception through headphones. These recordings will be made with a calibrated system that allows "audio-compliant" listening. The objective of these commented listening sessions is to get the participants to express their feelings about 15 to 20 sound samples of rail traffic representative of the variety of traffic on the pilot site and the instantaneous annoyance caused to the participants in the study, and to explain the acoustic characteristics of the train passages that are involved in this annoyance. The expected results are a better understanding of the perception of the different acoustic characteristics of railway noise events and the influence of these characteristics on annoyance.

## 5. DATA ANALYSIS

### 5.1 Analysis of noise measurement data

A detection of railway noise peaks will be carried out from the measured acoustic data. This will be done using an acoustic threshold crossing algorithm adapted to the distance of the measurement site from the railway tracks and the background noise of the area. This process will be completed by the information provided by the expert sensor with sound source detection, installed in the centre of the study area, which will make it possible to identify precisely the noise peaks due to rail traffic. Once these noise peaks have been identified, acoustic descriptors (LAmax, LAeq, SEL, duration, etc.) will be calculated for each train passage.

### 5.2 Use of interviews, questionnaires and notes of annoyance when trains pass

The semi-structured interviews will give rise to a content analysis that will allow a methodical and objective examination of the content of the interviews. These interviews will allow the identification of the psychosociological determinants of annoyance due to railway noise as well as the specific elements that these residents may complain about. The co-determinants of annoyance commonly identified in the literature are sensitivity to noise, visibility of the tracks, and the attitude of the subjects towards them. The interviews will also aim to ascertain whether there are any others. The answers provided by the participants in these qualitative interviews may be used to improve the general questionnaire as part of a larger study conducted on a national scale. The responses to the general questionnaire will also be analyzed to provide information about the participants and their housing.

The instantaneous annoyance notes produced by the remote controls will be associated with the railway sound events and the acoustic descriptors assessed at the front of participants' dwelling (acoustic factors measured at the nearest measuring station and corrected for the sound level differential between the measuring station and the resident's position determined by the modelling carried out for the pilot site).

### 5.3 Ranking of railway noise events

A first step will be to analyse the instantaneous annoyance scores per participant to find out the extent of the variability of the responses provided and to rank the railway noise events according to the associated instantaneous annoyance score. Statistical analysis tools (multivariate factor analysis: principal component analysis, discriminant analysis...) will then be used to identify the acoustic descriptors that best explain the variability of instantaneous annoyance. These analyses will be carried out for all participants and by groups of participants according to the rail noise zone in which they are located (moderate, intermediate, high), taking into account in particular the conditions at the time of scoring (indoors with the windows closed, indoors with the windows open or outdoors).

Once the acoustic factors influencing the instantaneous annoyance have been selected, the construction of an annoyance prediction model can be studied on the basis of these influencing factors and on the basis of statistical regression tools.

In addition, the two methods of assessing instantaneous annoyance used (remote control and commented listening to sound samples) could be compared in terms of acceptability to the participants in order to determine whether they give the same trends in the ranking of the instantaneous annoyance score expressed or not.

## 6. EXPECTED RESULTS

The expected results of this feasibility study are a methodology and assessment tools that can be implemented in a large-scale survey:

- A questionnaire to collect the perception of people living near railway infrastructures with regard to railway noise, in terms of the perception of noise from different categories of trains and the instantaneous annoyance caused by passing trains, as well as general information (information on the participants, the characteristics of their homes, their living environment, their perception about noise, their representation of the source, their professional situation, etc.).
- An electronic tool (remote control) to record levels of instantaneous annoyance experienced when different categories of trains pass.

This feasibility study will also make it possible to gather initial information on the ranking of trains according to the annoyance they cause and the weight of the various acoustic characteristics (intensity, duration, spectral content, suddenness, etc.) in the short-term annoyance experienced, as well as feedback on the various approaches implemented in terms of results and acceptability, with a view to making recommendations for the deployment of a study on a national scale.

## 7. ACKNOWLEDGMENTS

The partners would like to thank the ANSES for its assistance in the framework of the French National Research Program for Environmental and Occupational Health (ANSES-22-EST-182). They thank Patricia Champelovier of the AME-MODIS department of the Gustave Eiffel University for her help in developing the interview grid and the questionnaire. They would also like to thank the Municipality of Savigny-sur-Orge and the Grand Orly Seine-Bièvre public territorial establishment for their help in implementing this study on their territory. Finally, the partners would like to thank the inhabitants of Savigny-sur-Orge for their active participation and involvement in this survey.